\documentclass[aps,twocolumn,preprintnumbers,showpacs,nofootinbib]{revtex4}
\usepackage{amsmath,amssymb,bm,epsfig}

\newcommand\eps{\epsilon}
\renewcommand\d{\partial}
\newcommand\grad{\bm{\nabla}}
\newcommand\p{{\bm{p}}}
\newcommand\q{{\bm{q}}}
\renewcommand\k{{\bm{k}}}
\newcommand\ep{\varepsilon_\p}
\newcommand\eq{\varepsilon_\q}
\newcommand\ek{\varepsilon_\k}
\newcommand\Ep{E_\p}
\newcommand\+{\dagger}
\newcommand\eF{\varepsilon_{\mathrm{F}}}
\newcommand\Tr{\mathop{\mathrm{Tr}}}
\newcommand\eb{\varepsilon_\mathrm{b}}
\newcommand\Veff{V_\mathrm{eff}}
\newcommand\muB{{\mu_\mathrm{B}}}
\newcommand\wph{\omega_\mathrm{ph}}
\newcommand\fermi{f_\mathrm{F}}
\newcommand\bose{f_\mathrm{B}}
\newcommand\Tc{{T_\mathrm{c}}}

\def\Jvol<#1,#2,#3>{#1}
\def\Jpage<#1,#2,#3>{#2}
\def\Jyear<#1,#2,#3>{#3}
\newcommand\journal[1]{\textbf{\Jvol<#1>}, \Jpage<#1> (\Jyear<#1>)}
\newcommand\PRL[1]{Phys.\ Rev.\ Lett.\ \journal{#1}}
\newcommand\PRA[1]{Phys.\ Rev.\ A \journal{#1}}
\newcommand\PRB[1]{Phys.\ Rev.\ B \journal{#1}}
\newcommand\PRC[1]{Phys.\ Rev.\ C \journal{#1}}
\newcommand\PRD[1]{Phys.\ Rev.\ D \journal{#1}}

\newcommand\PLA[1]{Phys.\ Lett.\ A \journal{#1}}
\newcommand\PLB[1]{Phys.\ Lett.\ B \journal{#1}}

\newcommand\NJP[1]{New J.\ Phys.\ \journal{#1}}

\begin{document}
\preprint{INT-PUB 06-13, TKYNT-06-11}

\title{Unitary Fermi gas at finite temperature in the $\eps$ expansion}
\author{Yusuke~Nishida}
\email{nishida@nt.phys.s.u-tokyo.ac.jp}
%\homepage{http://tkynt2.phys.s.u-tokyo.ac.jp/~nishida/}
\affiliation{Department of Physics, University of Tokyo,
             Tokyo 113-0033, Japan}
\affiliation{Institute for Nuclear Theory, University of Washington,
             Seattle, Washington 98195-1550, USA}

\begin{abstract}
 Thermodynamics of the unitary Fermi gas at finite temperature is
 investigated from the perspective of the expansion over $\eps=4-d$ with
 $d$ being the dimensionality of space.  We show that the thermodynamics
 is dominated by bosonic excitations in the low temperature region
 $T\ll\Tc$.  Analytic formulas for the thermodynamic functions as
 functions of the temperature are derived to the lowest order in $\eps$
 in this region.  In the high temperature region where $T\sim\Tc$,
 bosonic and fermionic quasiparticles are excited.  We determine the
 critical temperature $\Tc$ of the superfluid phase transition and the
 thermodynamic functions around $\Tc$ to the leading and next-to-leading
 orders in $\eps$. 
\end{abstract}

\date{August 2006}

\maketitle

\section{Introduction}
The Fermi gas with zero-range interaction at infinite scattering
length~\cite{Eagles,Leggett,Nozieres}, frequently referred to 
as the unitary Fermi gas,
has attracted intense attention across many subfields of physics.
Experimentally, the system can be realized in atomic traps using the
Feshbach resonance and has been extensively 
studied~\cite{OHara,Jin,Grimm,Ketterle,Thomas,Salomon,Thomas-Tc}. 
Since the fermion density is the only dimensionful scale of
the unitary Fermi gas, its properties are universal, i.e.,
independent of details of the interparticle interaction.
The unitary Fermi gas is an idealization of dilute nuclear
matter and may be relevant to the physics of neutron
stars~\cite{Bertsch}.  It has been also suggested that its understanding 
may be important for the high-$\Tc$ superconductivity~\cite{highTc}. 

The austere simplicity of the unitary Fermi gas implies great
difficulties for theoretical treatments, because there seems to be no 
parameter for a perturbation theory.  
Recently, we have proposed a new approach for the unitary Fermi gas
based on the systematic expansion in terms of the dimensionality
of space~\cite{Nishida-Son1,Nishida-Son2}, utilizing the specialty of
\textit{four} or \textit{two} spatial dimensions in the unitarity 
limit~\cite{nussinov04}.  
In this approach, one would extend the problem to arbitrary spatial
dimensions $d$ and consider $d$ is close to four or two. 
Then one performs all calculations treating $\eps=4-d$ or $\bar\eps=d-2$
as a small parameter for the perturbative expansion.  Results for the
physical case of three spatial dimensions are obtained by extrapolating
the series expansions to $\eps\,(\bar\eps)=1$, or more appropriately, by
matching the two series expansions. 

We used this $\eps$ expansion around four spatial dimensions to
calculate thermodynamic functions and fermion quasiparticle spectrum in
the unitarity limit and found results which are quite
consistent with those obtained by Monte Carlo simulations and
experiments~\cite{Nishida-Son1,Nishida-Son2}.  This $\eps$ expansion has
been successfully applied to atom-dimer and dimer-dimer scatterings in
vacuum~\cite{Rupak-dimer}.  Thus there are compelling reasons to hope
that the limit $d\to4$ is not only theoretically interesting but also
practically useful, despite the fact that the expansion parameter $\eps$ 
is one at $d=3$.  Very recently, the phase structure of polarized Fermi
gas near the unitarity point has been investigated based on the $\eps$
expansion~\cite{Nishida-Son2,Rupak-polarized}. 

In this paper, we extend our approach to investigate the thermodynamics
of unitary Fermi gas at finite temperature $T$, below and above the
critical temperature $T=\Tc$ of the superfluid phase transition.  So
far, such a problem has been studied relying on the mean-field
description with
fluctuations~\cite{Melo,Haussmann,Holland,Timmermans,Ohashi,Milstein,Perali,Liu,Nishida-Abuki,Abuki,Zwerger},
the virial expansion~\cite{Ho,Horowitz,Rupak-finite-T}, or the Monte
Carlo
simulations~\cite{Wingate-Tc,bulgac-Tc,Lee-Schafer,burovski-Tc,Akkineni-Tc}. 
The main advantage of our approach is that $\eps=4-d$ provides the small
parameter of the perturbative expansion, and thus, analytic and
systematic study below and above $\Tc$ is possible. 
We also note that the critical dimension of a superfluid-normal phase
transition is also four, which makes weak-coupling calculations reliable
at any temperature for the small $\eps$. 

First, we review the $\eps$ expansion for the unitary Fermi gas around
four spatial dimensions in Sec.~\ref{sec:review}.  Then we discuss that
the thermodynamics in the low temperature region $T\ll\Tc$ is dominated
by bosonic excitations.  Analytic formulas for the thermodynamic
functions as functions of the temperature to the lowest order in $\eps$ 
are shown in Sec.~\ref{sec:below-Tc}. 
The behavior of the thermodynamic functions above $\Tc$ to the leading
and next-to-leading orders in $\eps$ is discussed in
Sec.~\ref{sec:above-Tc}.  In particular, we put emphasis on the
determination of the critical temperature $\Tc$ by matching the $\eps$
expansion with the expansion around two spatial dimensions. 
The thermodynamic functions at $\Tc$ are also calculated, which are 
compared to the results from recent Monte Carlo simulations. 
Finally, summary and concluding remarks are given in
Sec.~\ref{sec:summary}.  \newpage

\section{Review of $\eps$ expansion \label{sec:review}}
Here we briefly review the $\eps$ expansion for the unitary Fermi gas
around four spatial dimensions and quote some results at zero
temperature to the leading order in $\eps$ just for the convenience of
later discussions at finite temperature.  The detailed account of the 
$\eps$ expansion is found in Refs.~\cite{Nishida-Son1,Nishida-Son2}.

\subsection{Lagrangian and Feynman rules}
The extension to finite temperature $T$ follows from the prescription of
the imaginary time formalism.  The system under consideration is
described by the sum of following Lagrangian densities (here and below
$\hbar=1$ and $k_\mathrm{B}=1$): 
\begin{align}
 \begin{split}
  \mathcal{L}_0 & = \Psi^\+\left(\d_\tau - \frac{\sigma_3\grad^2}{2m}
  - \sigma_+\phi_0 - \sigma_-\phi_0\right)\Psi \\ 
  & \qquad + \varphi^*\left(\d_\tau - \frac{\grad^2}{4m}\right)\varphi
  + \frac{\phi_0^{\,2}}{c_0}\,,
 \end{split} \\ \notag\\
 \begin{split}
  \mathcal{L}_1 & = - g\Psi^\+\sigma_+\Psi\varphi 
  - g\Psi^\+\sigma_-\Psi\varphi^* - \mu\Psi^\+\sigma_3\Psi \\
  & \qquad - \left(2\mu-\frac{g^2}{c_0}\right)\varphi^*\varphi 
  + \frac{g\phi_0}{c_0}\varphi + \frac{g\phi_0}{c_0}\varphi^*\,, 
 \end{split} \\ \notag\\
 \mathcal{L}_2 & = -\varphi^*\left(\d_\tau
 - \frac{\grad^2}{4m}\right)\varphi + 2\mu\varphi^*\varphi\,.
\end{align}
The propagators of fermion and boson are generated by $\mathcal{L}_0$.
The fermion propagator $G$ is a $2\times2$ matrix, 
\begin{equation}
 G(i\omega_n,\p) = \frac1{(i\omega_n)^2-E_\p^{\,2}}
  \begin{pmatrix}
   i\omega_n + \ep & -\phi_0 \\
   -\phi_0 & i\omega_n-\ep
  \end{pmatrix},
\end{equation}
where $\ep=\p^2/2m$ is the kinetic energy of nonrelativistic particles
and $\phi_0$ chosen to be real is the condensate in the superfluid
ground state.  $E_\p=\sqrt{\ep^{\,2}+\phi_0^{\,2}}$ is the excitation
energy of the fermion quasiparticle.  The boson propagator $D$ is given
by
\begin{equation}
  D(i\nu_n,\p) = \left(i\nu_n - \frac{\ep}2\right)^{-1}.
\end{equation}
$\omega_n=2\pi T(n+\frac12)$ and $\nu_n=2\pi Tn$ are discrete Matsubara
frequencies for fermion and boson with an integer
$n=0,\pm1,\pm2,\cdots$.  

The unitary Fermi gas around four spatial dimensions is described by the
weakly interacting system of fermionic and bosonic quasiparticles, whose
coupling $g\sim\eps^{1/2}$ in $\mathcal{L}_1$ is given by 
\begin{equation}
 g = \frac{(8\pi^2\epsilon)^{1/2}}m
  \left(\frac{m\phi_0}{2\pi}\right)^{\epsilon/4}.
\end{equation}
The chemical potential $\mu\sim\eps\phi_0$ in $\mathcal{L}_1$ is treated
as a small perturbation in our formulation.  We define the boson
chemical potential as 
\begin{equation}
 \muB=2\mu-\frac{g^2}{c_0}.
\end{equation}
In the dimensional regularization we use, when $c_0$ is negative,
$-g^2/c_0\simeq\eb$ gives the binding energy of the boson to the leading
order in $\eps$.  Throughout this paper, we consider the vicinity of the
unitary point where $\eb\sim\eps\phi_0$. 

Finally the two additional vertices for the boson propagator in
$\mathcal{L}_2$ play a role of counter terms so as to avoid double
counting of certain types of diagrams which are already taken into
$\mathcal{L}_0$ and $\mathcal{L}_1$.  The first vertex in the momentum
space is given by 
\begin{equation}\label{eq:Pi_0}
  \Pi_0(p_0,\p) = p_0 -\frac{\ep}2.
\end{equation}

\subsection{Power counting rule of $\epsilon$  \label{sec:power-counting}} 

\begin{figure}[tp]
 \includegraphics[width=0.45\textwidth,clip]{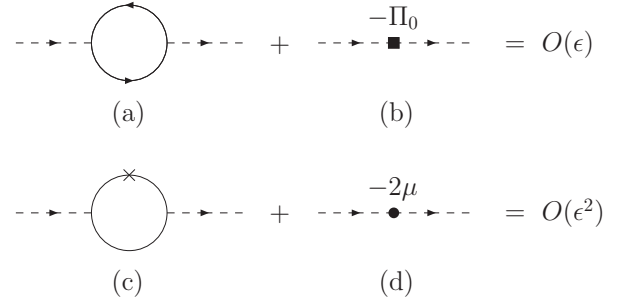}
 \caption{Restoration of naive $\epsilon$ counting for the boson
 self-energy.  The fermion loop in (c) goes around clockwise and
 counterclockwise.  Solid (dotted) lines represent the fermion (boson)
 propagator $-G$ $(-D)$, while the cross in (c) represents the $\mu$
 insertion to the fermion propagator.  \label{fig:cancel}} 
\end{figure}

The power counting rule of $\epsilon$ is summarized as follows. 
\begin{enumerate}
 \item We consider $\mu/\phi_0\sim\eps$ and regard $\phi_0$ as $O(1)$.
 \item For any Green's function, we write down all Feynman diagrams
       according to the Feynman rules using the propagators from
       $\mathcal{L}_0$ and the vertices from $\mathcal{L}_1$. 
 \item If there is any subdiagram of the type in Fig.~\ref{fig:cancel}(a) 
       or Fig.~\ref{fig:cancel}(c), we add the same Feynman diagram
       where the subdiagram is replaced by a vertex from
       $\mathcal{L}_2$, Fig.~\ref{fig:cancel}(b) or
       Fig.~\ref{fig:cancel}(d).
 \item The power of $\eps$ for the given Feynman diagram will be
       $O(\epsilon^{N_g/2+N_\mu})$, where $N_g$ is the number of
       couplings $g$ and $N_\mu$ is the number of chemical potential
       insertions.
 \item The only exception is the one-loop vacuum diagram of fermion with
       one $\mu$ insertion, which is $O(1)$ instead of the naive
       $O(\eps)$. 
\end{enumerate}
This power counting rule holds in the low temperature region where
the condensate is still large compared to the chemical potential 
$\mu/\phi_0\sim\eps$, while it breaks down near the critical temperature 
because $\phi_0\to0$ at $T\to\Tc$.  In the high temperature region
$T\sim\Tc$, a minor modification of the power counting rule is necessary 
as we discuss at the beginning of Sec.~\ref{sec:above-Tc}.

\subsection{Leading order results at $T=0$}
Here we show only lowest order results on the thermodynamic functions at
zero temperature, which is equivalent to the mean-field approximation.
The results up to the next-to-leading order in $\eps$ are found in
Refs.~\cite{Nishida-Son1,Nishida-Son2}.  The effective potential to the
leading order in $\eps$ is
\begin{equation}\label{eq:Veff}
 V_\mathrm{eff}(\phi_0) = \left[\frac{\phi_0}3-\frac{\muB}{2\eps}\right]
  \left(\frac{m\phi_0}{2\pi}\right)^{d/2},
\end{equation}
from which the condensate follows as
\begin{equation}\label{eq:phi_0}
 \phi_0 = \frac\muB\eps.
\end{equation}
Note that the previously made assumption $\mu/\phi_0=O(\epsilon)$ is
now checked.  Then the pressure at zero temperature is given by 
\begin{equation}\label{eq:P}
 P_0 = -\Veff(\phi_0) 
  = \frac{\phi_0}6\left(\frac{m\phi_0}{2\pi}\right)^{d/2}.
\end{equation}
From the fermion number density 
\begin{equation}\label{eq:N}
 N = \frac{\d P_0}{\d\mu} 
  = \frac1\epsilon\left(\frac{m\phi_0}{2\pi}\right)^{d/2},
\end{equation}
we define the Fermi energy through the relationship in the ideal Fermi
gas in $d$ spatial dimensions; 
\begin{equation}\label{eq:eF}
 \eF = \frac{2\pi}m
  \left[ \frac12\Gamma\left(\frac d2+1\right) N \right]^{2/d}
  = \frac{\phi_0}{\epsilon^{2/d}}.
\end{equation}
Then the energy density is given by
\begin{equation}
 E_0 = \mu N-P_0 = \left[\frac{\phi_0}3-\frac\eb{2\eps}\right]
  \left(\frac{m\phi_0}{2\pi}\right)^{d/2},
\end{equation}
where $\eb=-g^2/c_0>0$ is the binding energy of the boson.  
The chemical potential at the fixed Fermi energy $\eF$ is obtained from
Eqs.~(\ref{eq:phi_0}) and (\ref{eq:eF}) as 
\begin{equation}\label{eq:mu}
 \mu_0=\frac{\eps^{3/2}}2\eF-\frac\eb2.
\end{equation}

\section{Thermodynamics below $\Tc$ \label{sec:below-Tc}}
Now we investigate the thermodynamics of the Fermi gas at finite
temperature near the unitarity limit.  At zero temperature, we found that
there exist two difference energy scales in the system; the scale of the
condensate $\phi_0$ and that of the chemical potential
$\mu\sim\eps\phi_0\ll\phi_0$.  Accordingly, we can consider two
temperature regions where the unitary Fermi gas exhibits different
thermodynamics.

One is the low temperature region where $T\sim\eps\phi_0$.  In this
region, the energy gap of the fermion quasiparticle excitation
$\Delta\sim\phi_0$ is still large compared to the temperature.
Therefore, thermal excitations of the fermion quasiparticle are
exponentially suppressed by a factor $e^{-\Delta/T}\sim e^{-1/\eps}$.
The thermodynamics in this region is dominated by the bosonic phonon
excitations. 
The other temperature region is the high temperature region where
$T\sim\phi_0$.  ($\phi_0$ represents the condensate at zero
temperature.)  In this region, the condensate decreases and eventually
vanishes at the critical temperature $\Tc$.  Fermions and bosons are
equally excited here.  We defer our discussion on the high temperature
region to Sec.~\ref{sec:above-Tc} and concentrate on the thermodynamics
in the low temperature region $T\ll\Tc$ in this section.

\subsection{Phonon spectrum}

\begin{figure}[tp]
 \includegraphics[width=0.46\textwidth,clip]{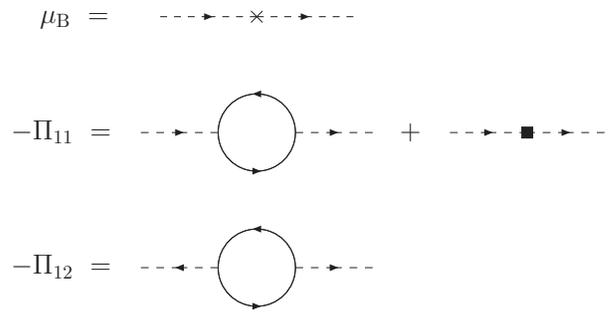}
 \caption{Boson's self-energy diagrams contributing to the order
 $O(\eps)$.  The vertex $\Pi_0$ from $\mathcal{L}_2$ needs to be added
 to the second diagram. The last diagram gives the off-diagonal part of
 the self-energy.  Other elements are given by
 $\Pi_{22}(p)=\Pi_{11}(-p)$ and $\Pi_{21}(p)=\Pi_{12}(p)^*$. 
 \label{fig:boson}}   
\end{figure}

The thermodynamics in the low temperature region $T\ll\phi_0$ is
dominated by the phonon excitations.  In order to determine the phonon
spectrum, we first study the boson self-energy at zero temperature.  To
the order of $O(\eps)$, there are three types of contributions to the
boson self-energy as depicted in Fig.~\ref{fig:boson}.  In addition to
the chemical potential insertion $\muB=2\mu+\eb$, the one-loop diagrams 
contribute to the diagonal part $\Pi_{11}(p)$ and off-diagonal part
$\Pi_{12}(p)$ of the boson self-energy.  The vertex $\Pi_0(p)$ from
$\mathcal{L}_2$ is necessary for $\Pi_{11}(p)$ according to the power
counting rule described in Sec.~\ref{sec:review}.  Other elements can be
obtained from $\Pi_{22}(p)=\Pi_{11}(-p)$ and
$\Pi_{21}(p)=\Pi_{12}(p)^*$.  

The diagonal part of the boson self-energy of $O(\eps)$ is given by the
sum 
\begin{equation}
 \Pi_{11}(p) = \Pi_0(p) + \Pi_\mathrm{a}(p),
\end{equation}
where $\Pi_0(p)$ is defined in Eq.~(\ref{eq:Pi_0}) and
$\Pi_\mathrm{a}(p)$ is given by
\begin{widetext}
\begin{equation}\label{eq:Pi_a}
 \begin{split}
  -\Pi_\mathrm{a}(p) %& = -g^2\int\!\frac{dk}{(2\pi)^{d+1}}
  %\Tr\left[G\left(k-\frac p2\right)
  %\sigma_+G\left(k+\frac p2\right)\sigma_-\right]
  &= g^2\int\!\frac{idk_0d\k}{(2\pi)^{d+1}}\, 
  G_{11}\!\left(k+\frac p2\right)G_{22}\!\left(k-\frac p2\right) \\ 
  &= g^2\int_\k \frac1{4E_{\k-\frac\p2}E_{\k+\frac\p2}} 
  \left[\frac{(E_{\k-\frac\p2}+\varepsilon_{\k-\frac\p2})
  (E_{\k+\frac\p2}+\varepsilon_{\k+\frac\p2})}
  {E_{\k-\frac\p2}+E_{\k+\frac\p2}-p_0}
  +\frac{(E_{\k-\frac\p2}-\varepsilon_{\k-\frac\p2})
  (E_{\k+\frac\p2}-\varepsilon_{\k+\frac\p2})}
  {E_{\k-\frac\p2}+E_{\k+\frac\p2}+p_0}\right].
 \end{split}
\end{equation}
\end{widetext}
Here we introduced the shorthand notation,
\begin{equation}
 \int_\k \equiv \frac{d\k}{(2\pi)^d}.
\end{equation}
Since we are interested in physics at the scale of temperature
$T\ll\phi_0$, it is sufficient to evaluate the self-energy when the
external momentum is small compared to the condensate 
$(p_0,\ep)\sim T\ll\phi_0$.  Expanding $\Pi_{11}(p)$ in terms of
$p/\phi_0$ and performing the $\k$ integration with the use of the
formula  
\begin{equation}\label{eq:formula}
 \int_0^\infty\!dz\,\frac{z^{\alpha-1}}{(z+1)^\beta}
  =\frac{\Gamma(\alpha)\Gamma(\beta-\alpha)}{\Gamma(\beta)},
\end{equation}
we obtain
\begin{equation}
 \Pi_{11}(0) = -g^2 \int_\k \frac{E_\k^{\,2}+\ek^{\,2}}{4E_\k^{\,3}}
  =\frac32\eps\phi_0+O(\eps^2).
\end{equation}

Similarly, the off-diagonal part of the boson-self energy of $O(\eps)$
is given by 
\begin{align}
 -\Pi_{12}(p)&=g^2\int\!\frac{idk_0d\k}{(2\pi)^{d+1}}\,
 G_{12}\left(k+\frac p2\right)G_{12}\left(k-\frac p2\right) \notag\\
 &=-g^2\int_\k 
 \frac{\phi_0^{\,2}}{4E_{\k-\frac\p2}E_{\k+\frac\p2}}\,\times\\
 &\left[\frac{1}{E_{\k-\frac\p2}+E_{\k+\frac\p2}-p_0}
 +\frac{1}{E_{\k-\frac\p2}+E_{\k+\frac\p2}+p_0}\right]. \notag
\end{align}
Expanding $\Pi_{12}(p)$ in terms of $p/\phi_0$ and performing the 
integration over $\k$, we obtain
\begin{equation}
 \Pi_{12}(0) = g^2 \int_\k \frac{\phi_0^{\,2}}{4E_\k^{\,3}}
  =\frac12\eps\phi_0+O(\eps^2).
\end{equation}

As a result of the resummation of these self-energies, the resummed 
boson propagator $\mathcal D$ is expressed by the following $2\times2$
matrix:  
\begin{equation}\label{eq:resum}
 \begin{split}
  & \mathcal D(p_0,\p) = \\ & \qquad
  \begin{pmatrix}
   D(p)^{-1}+\muB-\Pi_{11} & -\Pi_{12} \\
   -\Pi_{21} & D(-p)^{-1}+\muB-\Pi_{22}
  \end{pmatrix}^{-1},
 \end{split}
\end{equation}
where $\Pi_{22}(p)=\Pi_{11}(-p)$ and $\Pi_{21}(p)=\Pi_{12}(p)^*$.  Using
the self-energies calculated above, the dispersion relation of the boson
$\wph(\p)$ can be obtained by solving the equation
$\det[D^{-1}(\omega,\p)]=0$ in terms of $\omega$ as
\begin{equation}\label{eq:boson}
 \wph(\p)=\sqrt{\left(\frac\ep2-\muB+\eps\phi_0\right)
  \left(\frac\ep2-\muB+2\eps\phi_0\right)}.
\end{equation}
Note that this expression is valid as long as $\ep\ll\phi_0$ because of
the expansions made to evaluate the boson self-energies.  Substituting
the solution of the gap equation at zero temperature in
Eq.~(\ref{eq:phi_0}), $\muB=\eps\phi_0$, the phonon spectrum is
determined to be
\begin{equation}\label{eq:phonon}
 \wph(\p)=\sqrt{\frac\ep2\left(\frac\ep2+\eps\phi_0\right)}.
\end{equation}
For the small momentum $\ep\ll\eps\phi_0$, the dispersion relation 
becomes linear in the momentum as $\wph\simeq c_\mathrm{s}|\p|$, 
remaining gapless in accordance with the Nambu-Goldstone theorem. 
The sound velocity of phonon $c_\mathrm{s}$ is given by
\begin{equation}
 \begin{split}
  c_\mathrm{s}=\sqrt{\frac{\eps\phi_0}{4m}}+O(\eps^{3/2})
  \simeq v_\mathrm{F}\sqrt{\frac{\eps^{3/2}}8},
 \end{split}
\end{equation}
where $v_\mathrm{F}=(2\eF/m)^{1/2}$ is the Fermi velocity.  For the
large momentum $\ep\gg\eps\phi_0$, the dispersion relation approaches
that of the free boson as $\wph\simeq\ep/2$.

\subsection{Effective potential and condensate}

\begin{figure}[tp]
 \includegraphics[width=0.15\textwidth,clip]{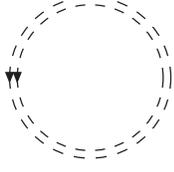}
 \caption{One-loop diagram of boson contributing to the effective
 potential at finite temperature. The dotted double line represents the
 resummed boson propagator $\mathcal D$ in Eq.~(\ref{eq:resum}).
 \label{fig:phonon}}
\end{figure}

At finite temperature, the phonon excitations contribute to the
effective potential, and consequently, the magnitude of the condensate 
decreases.  The temperature dependent part of the effective potential
$V_T(\phi_0)$ to the lowest order in $\eps$ is given by the one-loop
diagram of the boson with the resummed propagator in Eq.~(\ref{eq:resum})
(Fig.~\ref{fig:phonon}): 
\begin{equation}\label{eq:V_T}
 \begin{split}
  V_T(\phi_0) &=\frac T2\sum_n\int_\p
  \Tr\ln\left[\mathcal D(i\nu_n,\p)^{-1}\right]\\
  &=T\int_\p \ln\left[1-e^{-\wph(\p)/T}\right].
 \end{split}
\end{equation}
The $n$-point interaction vertex among phonons $\varphi^n$ is of the
order $\eps^{n/2}$ and appears in the effective potential only at higher
orders.  Then the contribution of $V_T(\phi_0)$ to the gap equation is
\begin{equation}\label{eq:gap_eq}
 \frac{\d V_T(\phi_0)}{\d\phi_0}
  =\int_\p \bose(\wph)\,\frac{\d\wph}{\d\phi_0},
\end{equation}
where $\bose(x)=1/(e^{x/T}-1)$ is the Bose distribution function and 
$\d\wph/\d\phi_0$ is given from Eq.~(\ref{eq:boson}) by
\begin{equation}
 \frac{\d\wph}{\d\phi_0}=\eps\frac{3\ep+2\eps\phi_0}{4\wph}.
\end{equation}

There are two limiting cases where the integration over $\p$ in
Eq.~(\ref{eq:gap_eq}) can be analytically performed.  Since the integral 
is dominated by the integration region where $\ep\sim T$, we can
approximate the phonon spectrum by its linear branch 
$\wph(\p)\simeq c_\mathrm{s}|\p|$ when the temperature is very low
$T\ll\eps\phi_0$.  In this case, the integration over $\p$ at $d=4$
leads to
\begin{equation}
 \frac{\d V_T(\phi_0)}{\d\phi_0}\simeq
  \frac{8\zeta(3)T}{\phi_0}\left(\frac{mT}{2\pi}\right)^2.
\end{equation}
On the other hand, when the temperature is located in the intermediate
region $\eps\phi_0\ll T\ll\phi_0$, the phonon spectrum can be
approximated by its quadratic branch $\wph\simeq\ep/2$.  In this case,
the integration over $\p$ in Eq.~(\ref{eq:gap_eq}) at $d=4$ results in
\begin{equation}
 \frac{\d V_T(\phi_0)}{\d\phi_0}\simeq\eps\frac{(mT)^2}4.
\end{equation}

Now, from the gap equation $\d[\Veff(\phi)+V_T(\phi)]/\d\phi=0$ with
$\phi\equiv\phi_0+\phi_T$, one finds the temperature dependent
correction of the condensate $\phi_T$ satisfies
\begin{equation}
 \frac{\d V_\mathrm{eff}(\phi_0)^2}{\d\phi_0^{\,2}}\phi_T
  +\frac{\d V_T(\phi_0)}{\d\phi_0}=0,
\end{equation}
where $\Veff(\phi_0)$ is the effective potential at zero temperature in
Eq.~(\ref{eq:Veff}).  To the leading order in $\eps$, $\phi_T$ at
$T\ll\eps\phi_0$ is given by
\begin{equation}
 \phi_T=-\frac{8\zeta(3)T^3}{\phi_0^{\,2}},
\end{equation}
while at $\eps\phi_0\ll T\ll\phi_0$,
\begin{equation}
 \phi_T=-\eps\frac{\pi^2T^2}{\phi_0}. 
\end{equation}
The condensate in total is $\phi=\phi_0+\phi_T$, which decreases as
the temperature increases.  Note that since $\phi_T\ll\eps\phi_0$, the
leading part of the condensate does not change in the temperature region
considered here $T\ll\phi_0$.  The effective potential is given by the
sum of the zero temperature and finite temperature parts; 
$\Veff(\phi_0+\phi_T)+V_T(\phi_0+\phi_T)\simeq\Veff(\phi_0)+V_T(\phi_0)$.

\subsection{Thermodynamic functions at low temperature}
The temperature dependent part of the pressure $P_\mathrm{ph}$ in the
low temperature region $T\ll\phi_0$ is given from the effective potential
in Eq.~(\ref{eq:V_T}) by
\begin{equation}\label{eq:P_ph}
 P_\mathrm{ph} = -V_T(\phi_0)
  = -T\int_\p \ln\left[1-e^{-\wph(\p)/T}\right].
\end{equation}
The phonon contributions to the fermion number density, the
entropy density, and the energy density are respectively computed from
the thermodynamic relations, 
$N_\mathrm{ph}=\d P_\mathrm{ph}/\d\mu$, 
$S_\mathrm{ph}=\d P_\mathrm{ph}/\d T$, and 
$E_\mathrm{ph}=\mu N_\mathrm{ph}+TS_\mathrm{ph}-P_\mathrm{ph}$.  Here we
show analytic expressions for these thermodynamic functions in the two
cases where the analytic evaluation of the $\p$ integration in
Eq.~(\ref{eq:P_ph}) is available.

When the temperature is very low $T\ll\eps\phi_0$, only the linear
branch of the phonon spectrum 
$\omega_\mathrm{ph}(\p)\simeq c_\mathrm{s}|\p|$ is important to the
thermodynamic functions.  In this case, the integration over $\p$ at
$d=4$ can be performed analytically to lead to
\begin{equation}
 P_\mathrm{ph}\simeq
  \frac{12\pi^2\zeta(5)}{(2\pi)^4}\frac{T^5}{c_\mathrm{s}^{\,4}}
  =\frac{12\zeta(5)}{\pi^2}\frac{m^2T^5}{(\eps\phi_0)^2}.
\end{equation}
Note that the exponent of $T$ in general spatial dimension $d$ is
$P_\mathrm{ph}\sim T^{d+1}$.  Accordingly, we obtain the phonon
contributions to the fermion number density, the entropy density, and
the energy density:
\begin{align}
 N_\mathrm{ph}&=-\frac{48\zeta(5)}{\pi^2}\frac{m^2T^5}{(\eps\phi_0)^3},
 \label{eq:N_ph} \\
 S_\mathrm{ph}&=\frac{60\zeta(5)}{\pi^2}\frac{m^2T^4}{(\eps\phi_0)^2},
 \phantom{\frac{\frac\int\int}{\frac\int\int}}\hspace{-4.5mm} \\
 E_\mathrm{ph}&=\frac{24\zeta(5)}{\pi^2}\frac{m^2T^5}{(\eps\phi_0)^2}
 \left(1+\frac\eb{\eps\phi_0}\right).
\end{align}

Since actual experiments or simulations are performed with the fixed
fermion density, it is useful to show the thermodynamic functions at
fixed $N$ instead of fixed $\mu$.  From Eqs.~(\ref{eq:phi_0}),
(\ref{eq:N}), and (\ref{eq:N_ph}), we find the chemical potential for
the fixed fermion density increases as a function of the temperature as 
\begin{equation}
 \mu=\mu_0+48\zeta(5)\frac{T^5}{\eps\phi_0^{\,4}}, 
\end{equation}
where $\mu_0$ represents the chemical potential at zero temperature in
Eq.~(\ref{eq:mu}).  Normalizing $\mu$ by the Fermi energy in
Eq.~(\ref{eq:eF}), we have 
\begin{equation}\label{eq:mu-N1}
 \frac\mu\eF =\frac{\mu_0}\eF
 +\frac{3\zeta(5)}{2\eps^3}\left(\frac{2T}\eF\right)^5.
\end{equation}
The other thermodynamic functions for the fixed fermion number density 
are given by 
\begin{align}
 \frac{P}{\eF N} &=\frac{P_0}{\eF N}
 +\frac{3\zeta(5)}{\eps^3}\left(\frac{2T}\eF\right)^5,\\
 \frac{E}{\eF N} &=\frac{E_0}{\eF N}
 +\frac{6\zeta(5)}{\eps^3}\left(\frac{2T}\eF\right)^5,
 \phantom{\frac{\frac\int\int}{\frac\int\int}}\hspace{-4.5mm} \\
 \frac{S}{N} &=\frac{15\zeta(5)}{\eps^3}\left(\frac{2T}\eF\right)^4.
 \label{eq:S-N1}
\end{align}
These expressions are valid in the low temperature region where
$T\ll\eps\phi_0$ and hence $T/\eF\ll\eps^{3/2}$.  The exponent of $T$ is
different from that in three spatial dimensions because we are expanding
around four spatial dimensions.  The correct exponent at $d=3$ is
recovered if we resum logarithmic corrections 
$\sim\left(-\eps\ln T\right)^{n}$ to infinite orders, which are formally
higher orders in $\eps$ and are not shown in the formulas above.

On the other hand, when the temperature is located in the intermediate
region $\eps\phi_0\ll T\ll\phi_0$, we can expand the phonon spectrum
$\omega_\mathrm{ph}(\p)$ in terms of $\eps\phi_0/\ep$ up to its first
order
\begin{equation}
 \wph(\p)\simeq\frac\ep2+\frac{\eps\phi_0}2. 
\end{equation}
In this case, the integration over $\p$ in Eq.~(\ref{eq:P_ph}) at $d=4$
can be performed analytically again to result in
\begin{equation}
 P_\mathrm{ph}\simeq\frac{\zeta(3)}{\pi^2}m^2T^3
  -\frac{m^2T^2}{12}\eps\phi_0.
\end{equation}
Note that the exponent of $T$ in general spatial dimension $d$ is
$P_\mathrm{ph}\sim T^{d/2+1}+T^{d/2}$.  Accordingly, we obtain the
temperature dependent parts of the fermion number density, the entropy
density, and the energy density:
\begin{align}
 N_\mathrm{ph}&=-\frac{m^2T^2}{6},\label{eq:N_ph2}\\
 S_\mathrm{ph}&=\frac{3\zeta(3)}{\pi^2}m^2T^2-\frac{m^2T}{6}\eps\phi_0,
 \phantom{\frac{\int}{\int}}\hspace{-4.5mm} \\
 E_\mathrm{ph}&=\frac{2\zeta(3)}{\pi^2}m^2T^3
 -\frac{m^2T^2}{6}\eps\phi_0\left(1-\frac\eb{2\eps\phi_0}\right).
\end{align}

From Eqs.~(\ref{eq:phi_0}), (\ref{eq:N}), and (\ref{eq:N_ph2}), we find
the chemical potential for the fixed fermion density increases as a
function of the temperature as
\begin{equation}
  \mu=\mu_0+\eps^{2}\frac{\pi^2}6\frac{T^2}{\phi_0}.
\end{equation}
Normalizing $\mu$ by the Fermi energy in Eq.~(\ref{eq:eF}), we have 
\begin{equation}\label{eq:mu-N2}
 \frac\mu\eF=\frac{\mu_0}\eF
 +\eps^{3/2}\frac{\pi^2}6\left(\frac T\eF\right)^2. 
\end{equation}
The other thermodynamic functions for the fixed fermion number density 
are given by 
\begin{align}
 \frac{P}{\eF N} &=\frac{P_0}{\eF N}+4\zeta(3)\left(\frac T\eF\right)^3
 -\eps^{3/2}\frac{\pi^2}6\left(\frac T\eF\right)^2,\\
 \frac{E}{\eF N} &=\frac{E_0}{\eF N}+8\zeta(3)\left(\frac T\eF\right)^3
 -\eps^{3/2}\frac{\pi^2}3\left(\frac T\eF\right)^2,
 \phantom{\frac{\frac\int\int}{\frac\int\int}}\hspace{-4.5mm} \\
 \frac{S}{N} &=12\zeta(3)\left(\frac T\eF\right)^2
 -\eps^{3/2}\frac{2\pi^2}3\frac T\eF.  \label{eq:S-N2}
\end{align}
These expressions are valid in the intermediate temperature region where
$\eps\phi_0\ll T\ll\phi_0$ and hence 
$\eps^{3/2}\ll T/\eF\ll\eps^{1/2}$.  The exponent of $T$ is different
from that in three spatial dimensions because we are expanding around
four spatial dimensions.  The correct exponent at $d=3$ is recovered if
we resum logarithmic corrections $\sim\left(-\frac\eps2\ln T\right)^{n}$
to infinite orders, which are formally higher orders in $\eps$ and are
not shown in the formulas above.

\subsection{Effective potential near $\Tc$}
At the end of this section, we study the behavior of the condensate
$\phi$ as a function of the temperature for a given $\mu$.  The
effective potential to the leading order in $\eps$ is given by one-loop
diagrams of fermion with and without one $\mu$ insertion, which is
equivalent to the mean-field approximation.  Since the critical
dimension of the superfluid-normal phase transition is four, the
mean-field approximation remains as a leading part at any temperature in
the limit $d\to 4$.  Then the leading contribution to the effective
potential at finite temperature is given by
\begin{equation}
  \Veff(\phi) = -\frac{\eb}{g^2}\phi^{2}  -\int_\p
 \left[E_\p-\frac{\ep}{\Ep}\mu+2T\ln\left(1{+}e^{-\Ep/T}\right)
 %+\fermi(\Ep)\frac{2\ep}{\Ep}\mu
 \right], 
\end{equation}
where $\fermi(x)=1/(e^{x/T}+1)$ is the Fermi distribution function.

\begin{figure}[tp]
 \includegraphics[width=0.45\textwidth,clip]{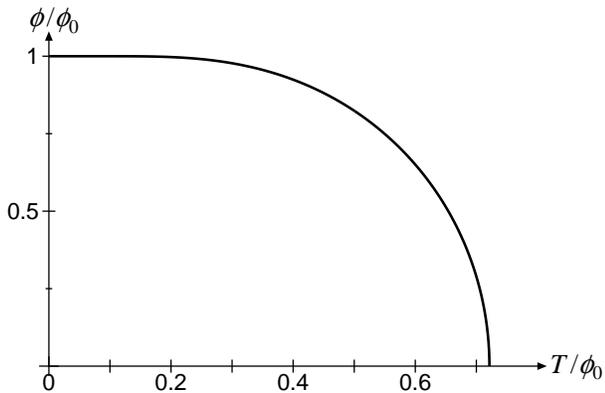}
 \caption{The behavior of the condensate $\phi$ as a function of the 
 temperature $T$ to the lowest order in $\eps$.  $\phi_0=\muB/\eps$ is 
 the value of the condensate at $T=0$ and the critical temperature is 
 located at $\Tc/\phi_0=1/(2\ln2)\approx0.721348$.
 \label{fig:mean-field}}  
\end{figure}

For the low temperature $T\ll\phi$, we can neglect the exponentially
small factor $e^{-\Ep/T}\ll1$ and the integration over $\p$ reproduces 
the effective potential at zero temperature in Eq.~(\ref{eq:Veff}).  In
the opposite limit where $\phi\ll T$, we can expand $\Veff(\phi)$ in
terms of $\phi/T$ to lead to  
\begin{align}
 &\Veff(\phi) - \Veff(0)\\ &\quad
 =\left[T\ln2-\frac\muB{2\eps}\right]\left(\frac{m\phi}{2\pi}\right)^2
 +\frac{\phi^{2}}{16T}\left(\frac{m\phi}{2\pi}\right)^2+\cdots,
 \notag
\end{align}
where $\muB=2\mu+\eb$. 
From the coefficient of the quadratic term in $\phi$, we can read the
critical temperature $\Tc$ of the second order phase transition to the
leading order in $\eps$ as
\begin{equation}\label{eq:Tc-muB}
 \Tc=\frac\muB{\eps\,2\ln2}+O(\eps),
\end{equation}
and the value of the condensate $\phi$ just below $\Tc$ as
\begin{equation}
 \phi^{2}=8T\left(\Tc-T\right)\ln2+O(\eps). 
\end{equation}
The condensate $\phi$ in the intermediate range of the temperature is
obtained by solving the gap equation $\d\Veff/\d\phi=0$:
\begin{equation}
 \phi-\frac\muB\eps+\int\!d\ep\frac{2\ep}\Ep\fermi(\Ep)=0.
\end{equation}
The numerical solution of the gap equation as a function of $T$ is shown
in Fig.~\ref{fig:mean-field}. 

Finally we note that critical exponents for an $\mathrm{O}(2)$-symmetric
theory will be in principle recovered if we resum logarithmic
corrections $\sim\left(\eps^m\ln|T-\Tc|\right)^n$ appearing at higher
orders due to the infrared physics of bosons with a zero Matsubara
frequency.  [We can see such a logarithmic correction, e.g., in
Eq.~(\ref{eq:NB_regular}).]   Resumming these logarithmic corrections to
infinite orders, the critical exponent shifts from its mean-field value
by $|T-\Tc|^{\#\eps^m}$ to produce the known results for the $XY$
universality class in the usual $\eps$ expansion.  In other words, we
cannot see the shift of the critical exponent at finite orders in our
$\eps$ expansion.

\section{Thermodynamics above $\Tc$ \label{sec:above-Tc}}

\subsection{Power counting rule of $\eps$ near $\Tc$} 
In the $\eps$ expansion at zero or low temperature, the chemical
potential is small compared to the condensate $\mu\sim\eps\phi_0$ and we
made expansions in terms of $\mu/\phi_0\sim\eps$ as well as the small
coupling $g\sim\eps^{1/2}$.  Near the critical temperature, the ratio
$\mu/\phi$ is no longer small because the condensate $\phi$ vanishes at
$T=\Tc$, while $\mu/\Tc$ is $O(\eps)$ as it is clear from
$\Tc=\muB/(\eps\,2\ln2)$ in Eq.~(\ref{eq:Tc-muB}).  Therefore, we can
still treat the chemical potential as a small perturbation near $\Tc$
and the same power counting rule of $\eps$ described in
Sec.~\ref{sec:power-counting} holds even above $\Tc$ just by replacing
$\phi_0$ with $T$.  Hereafter we regard $T\sim\Tc$ as $O(1)$.

\subsection{Boson's thermal mass}
First we study the self-energy of boson at $T\geq\Tc$.  The leading 
contribution to the self-energy is the chemical potential insertion
$\muB$ as well as the one-loop diagram $\Pi_{11}$ shown in
Fig.~\ref{fig:boson}: 
\begin{align}
 &\Pi_{11}(i\nu,\p)-\Pi_0(i\nu,\p) \notag\\
 &\quad=g^2\,T\sum_n\int_\k\,
 G_{11}(i\omega_n+i\nu,\k+\p)G_{22}(i\omega_n,\k) \notag\\
 &\quad=-g^2\int_\k \frac{1-\fermi(\varepsilon_{\k-\p/2})
 -\fermi(\varepsilon_{\k+\p/2})}{2\ek-i\nu+\ep/2}.
\end{align}
For the zero Matsubara frequency mode $\nu_n=0$ at the small momentum 
$\ep\ll T$, we have 
\begin{equation}\label{eq:thermal}
 \Pi_{11}(0,\bm{0}) = g^2\int_\k \frac{\fermi(\ek)}{\ek} 
  =\eps\,T\,2\ln2 +O(\eps^2).
\end{equation}
Therefore, the zero Matsubara frequency mode has the non-negative
thermal mass $\Pi_T\equiv\eps\,T\,2\ln2-\muB\sim O(\eps)$ at $T\geq\Tc$.
The condition of the vanishing thermal mass $\Pi_T=0$ gives the critical
temperature $\Tc=\muB/(\eps\,2\ln2)+O(\eps)$ equivalent to
Eq.~(\ref{eq:Tc-muB}).  As we will see below, at a sufficiently high
order in the perturbation theory near $\Tc$ [$O(\eps^2)$ or $O(\eps)$
compared to the leading term in the pressure or fermion density], a
resummation of the boson self-energy is needed to avoid infrared
singularities appearing in the zero Matsubara frequency mode.

\onecolumngrid
\subsection{Pressure}

\begin{figure}[tp]
 \includegraphics[width=0.45\textwidth,clip]{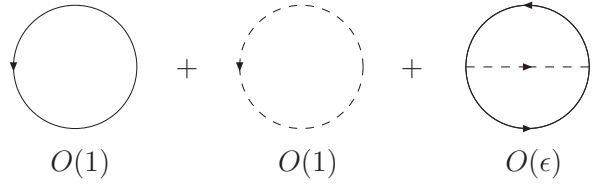}
 \caption{Three types of diagrams contributing to the pressure up to the 
 next-to-leading order in $\eps$.  Each $\mu$ $(\muB)$ insertion to the 
 fermion (boson) line reduces the power of $\eps$ by one. 
 \label{fig:potential}} 
\end{figure}

Now we calculate the thermodynamic functions at $T\geq\Tc$ to the
leading and next-to-leading orders in $\eps$.  There are three types of
diagrams contributing to the pressure up to the next-to-leading order in 
$\eps$ as depicted in Fig.~\ref{fig:potential}; one-loop diagrams of
fermion (boson) with and without one $\mu$ $(\muB)$ insertion and
two-loop diagram with a boson exchange.  Note that at $T\geq\Tc$, the
boson's one-loop diagram contributes as $O(1)$ as well as the fermion's
one-loop diagram.  Then the pressure from the one-loop diagrams is given
by 
\begin{equation}\label{eq:P1}
 \begin{split}
  P_1&=\int_\p
  \left[2T\ln\left(1+e^{-\ep/T}\right)-T\ln\left(1-e^{-\ep/2T}\right)
  +2\mu\fermi(\ep)+\muB\bose(\ep/2)\right]\\
  &= T\left[\frac{11}2\zeta(3)-\frac{9\ln2\,\zeta(3)+11\zeta'(3)}4\eps 
  +\frac{\pi^2}6\frac\mu T+\frac{2\pi^2}3\frac\muB T\right]
  \left(\frac{mT}{2\pi}\right)^{d/2}. 
 \end{split}
\end{equation}
The contribution from the two-loop diagram to the pressure, which is
$O(\eps)$, is given by
\begin{equation}
 \begin{split}
  P_2 & = g^2\,T^2\sum_{n,m}\int_{\p\q}
  G_{11}(i\omega_n,\p)G_{22}(i\omega_m,\q)D(i\omega_n-i\omega_m,\p-\q)\\ 
  &=-g^2\int_{\p\q}
  \frac{\fermi(\ep)\fermi(\eq)+\left[\fermi(\ep)+\fermi(\eq)\right]
  \bose(\varepsilon_{\p-\q}/2)}{\ep+\eq-\varepsilon_{\p-\q}/2}.
 \end{split}
\end{equation}
The numerical integrations over $\p$ and $\q$ result in
\begin{equation}\label{eq:P2}
  P_2 = -C_P\eps\left(\frac{mT}{2\pi}\right)^{d/2}T,
\end{equation}
where $C_P\approx8.4144$ is a numerical constant.  From
Eqs.~(\ref{eq:P1}) and (\ref{eq:P2}), we obtain the pressure up to the
next-to-leading order in $\eps$ as
\begin{equation}\label{eq:P-Tc}
 \begin{split}
  P = P_1+P_2 = T\left[\frac{11}2\zeta(3)-\frac{9\ln2\,\zeta(3) 
  +11\zeta'(3)}4\eps -\eps\,C_P +\frac{\pi^2}6\frac\mu T
  +\frac{2\pi^2}3\frac\muB T\right]\left(\frac{mT}{2\pi}\right)^{d/2}. 
 \end{split}
\end{equation}
The entropy density $S$ and the energy density $E$ to the same order can
be computed from the thermodynamic relations $S=\d P/\d T$ and 
$E=\mu N+TS-P$.

\subsection{Fermion number density}
The fermion number density to the next-to-leading order in $\eps$ cannot
be obtained simply by differentiating the pressure in
Eq.~(\ref{eq:P-Tc}) with respect to the chemical potential
$N=\d P/\d\mu$.  Since the pressure to the leading order in $\eps$ does
not depend on $\mu$ and the $\mu$ derivative $\d/\d\mu\sim1/\eps$
enhances the power of $\eps$ by one, we need to compute the one-loop 
diagrams with two $\mu$ $(\muB)$ insertions and the two-loop diagrams
with one $\mu$ $(\muB)$ insertion.  The fermion density from the
fermion's one-loop diagrams is given by
\begin{align}\label{eq:NF}
 \begin{split}
  N_{\mathrm{F}}&=2\int_\p
  \left[\fermi(\ep)+\frac{\mu}T\fermi(\ep)\fermi(-\ep)\right]\\
  &=\left[\frac{\pi^2}6-\frac{\pi^2\ln2+6\zeta'(2)}{12}\eps
  +\frac{2\ln2}T\mu\right]\left(\frac{mT}{2\pi}\right)^{d/2}. 
 \end{split}
\end{align}

\begin{figure}[tp]
 \includegraphics[width=0.2\textwidth,clip]{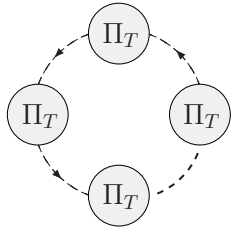}
 \caption{A type of diagrams resummed to resolve the infrared
 singularity in Eq.~(\ref{eq:infrared}).  \label{fig:resummation}}
\end{figure}

On the other hand, the boson's one-loop diagrams contribute to the
fermion density as
\begin{equation}\label{eq:infrared}
 N_{\mathrm{B}}=2\int_\p
 \left[\bose(\ep/2)-\frac{\muB}T\bose(\ep/2)\bose(-\ep/2)\right]. 
\end{equation}
Apparently, the last term has an infrared singularity because the Bose
distribution function behaves as $\bose(\ep/2)\simeq2T/\ep$ at small
momentum $\ep\ll T$.  In order to resolve this infrared singularity, the
resummation of the boson self-energy $\Pi_T=\Pi_{11}-\muB\sim O(\eps)$
is needed at the small momentum region $\ep\sim\mu$.  After resumming a
type of diagrams shown in Fig.~\ref{fig:resummation}, we can rewrite
$N_{\mathrm{B}}$ within the accuracy we are working as
\begin{align}\label{eq:NB}
 N_{\mathrm{B}}=2\int_\p \left[\bose(\ep/2+\Pi_T)
 -\eps\,2\ln2\,\bose(\ep/2)\bose(-\ep/2)\right]+O(\eps^2),
\end{align}
where we used $\Pi_{11}/T=\eps\,2\ln2$.  Now the first term is infrared
finite, where the boson's thermal mass $\Pi_T$ plays a role of an
infrared cutoff.  Integrating over $\p$ and expanding up to the
next-to-leading order in $\eps$, we have 
\begin{equation}\label{eq:NB_regular}
 2\int_\p \bose(\ep/2+\Pi_T)
  =\left[\frac{4\pi^2}3-\frac{2\pi^2\ln2+12\zeta'(2)}3\eps
    -8\left(1-\ln\frac{\Pi_T}T\right)\frac{\Pi_T}T+O(\eps^2)\right]
  \left(\frac{mT}{2\pi}\right)^{d/2}. 
\end{equation}
The logarithmic term $\sim\ln\Pi_T/T$ appears as a consequence of the
resummation.  The second term in Eq.~(\ref{eq:NB}), which is still
infrared divergent, will cancel with the infrared singularity existing
in the two-loop diagram.

The contribution from the two-loop diagrams to the fermion density is 
given by
\begin{equation}
 N_2 = -g^2\int_{\p\k}
  \frac{\fermi(\varepsilon_{\k+\p/2})\fermi(-\varepsilon_{\k+\p/2})
  \left[\fermi(\varepsilon_{\k-\p/2})+\bose(\ep/2)\right]
  -2\fermi(\varepsilon_{\k+\p/2})\bose(\ep/2)\bose(-\ep/2)}{\ek T}.
\end{equation}
The second term in the numerator contains the infrared singularity at
small $\ep$.  Extracting the divergent part, we can rewrite $N_2$ as
\begin{align}
 N_2 &= -g^2\int_{\p\k}
 \frac{\fermi(\varepsilon_{\k+\p/2})\fermi(-\varepsilon_{\k+\p/2})
 \left[\fermi(\varepsilon_{\k-\p/2})+\bose(\ep/2)\right]
 -2\left[\fermi(\varepsilon_{\k+\p/2})-\fermi(\ek)\right]
 \bose(\ep/2)\bose(-\ep/2)}{\ek T} \notag\\
 &\qquad +2g^2\int_{\p\k} 
 \frac{\fermi(\ek)}{\ek T} \bose(\ep/2)\bose(-\ep/2).
\end{align}
Now the first term is infrared finite.  One finds the $\k$ integration
in the second term can be performed to lead to $\Pi_{11}=\eps\,T\,2\ln2$
in Eq.~(\ref{eq:thermal}), which exactly cancels out the infrared
divergent part in Eq.~(\ref{eq:NB}).  The numerical integrations over
$\k$ and $\p$ in the first term result in 
\begin{equation}\label{eq:N2}
 N_2=-C_N\eps\left(\frac{mT}{2\pi}\right)^{d/2}
  +\eps\,4\ln2\int_\p \bose(\ep/2)\bose(-\ep/2),
\end{equation}
where $C_N\approx1.92181$ is a numerical constant.  

Gathering up all contributions, Eqs.~(\ref{eq:NF}), (\ref{eq:NB}), and
(\ref{eq:N2}), the fermion number density to the leading and
next-to-leading orders in $\eps$ is given by
\begin{align}\label{eq:N-Tc}
 \begin{split}
  N &= N_{\mathrm{F}}+N_{\mathrm{B}}+N_2\\
  &=\left[\frac{3\pi^2}2-\frac{3\pi^2\ln2+18\zeta'(2)}4\eps-\eps\,C_N
  +\frac{2\ln2}T\mu-8\left(1-\ln\frac{\Pi_T}T\right)\frac{\Pi_T}T
  \right]\left(\frac{mT}{2\pi}\right)^{d/2}. 
 \end{split}
\end{align}
We define the Fermi energy $\eF$ through the relationship in
Eq.~(\ref{eq:eF}) as 
\begin{equation}\label{eq:eF-Tc}
 \begin{split}
  \frac\eF{T}&=\sqrt{\frac{3\pi^2}2}
  \left[1+\left\{\frac{2\gamma-3-2\ln2}8-\frac{3\zeta'(2)}{2\pi^2}
  -\frac{C_N}{3\pi^2}+\frac18\ln\!\left(\frac{3\pi^2}2\right)\right\}\eps
  +\frac{2\ln2}{3\pi^2}\frac\mu{T}
  -\frac8{3\pi^2}\left(1-\ln\frac{\Pi_T}T\right)\frac{\Pi_T}T\right].
 \end{split}
\end{equation}
The logarithmic correction $(\Pi_T/T)\ln\Pi_T/T\sim\eps\ln\eps$ is a
consequence of the resummation in the infrared physics of bosons with
the zero Matsubara frequency, while it vanishes just at the critical
temperature $\Pi_\Tc=\eps\,\Tc\,2\ln2-\muB=0$.

\subsection{Critical temperature}\twocolumngrid
The critical temperature in units of the Fermi energy directly follows
from Eq.~(\ref{eq:eF-Tc}) with the use of the relationship
$2\mu+\eb=\eps\,\Tc\,2\ln2$:
%\begin{align}\label{eq:Tc}
% \frac\Tc\eF &= \sqrt{\frac2{3\pi^2}} \notag
% \left[1-\left\{\frac{2\gamma-3-2\ln2}8-\frac{3\zeta'(2)}{2\pi^2}
% \right.\right.\\  &\qquad \left.\left. -\frac{C_N-2(\ln2)^2}{3\pi^2} 
% +\frac18\ln\!\left(\frac{3\pi^2}2\right)\right\}\eps\right] \notag
% +\frac{\ln2}{3\pi^2}\frac\eb\eF \\
% &=0.260-0.0112\,\eps+0.0234\,\frac\eb\eF+O(\eps^2),
% \phantom{\frac{\frac\int\ }\ }\hspace{-4.5mm}
% %0.259899 -0.0111548 +0.0234102
%\end{align}
\begin{widetext}
\begin{equation}\label{eq:Tc}
 \begin{split}
  \frac\Tc\eF &= \sqrt{\frac2{3\pi^2}}
  \left[1-\left\{\frac{2\gamma-3-2\ln2}8-\frac{3\zeta'(2)}{2\pi^2}
  -\frac{C_N-2(\ln2)^2}{3\pi^2}
  +\frac18\ln\!\left(\frac{3\pi^2}2\right)\right\}\eps\right] 
  +\frac{\ln2}{3\pi^2}\frac\eb\eF \\
  &=0.260-0.0112\,\eps+0.0234\,\frac\eb\eF+O(\eps^2),
  \phantom{\frac{\frac\int\ }\ }  %0.259899 -0.0111548 +0.0234102
 \end{split}
\end{equation}
\end{widetext}
where the numerical value $C_N\approx1.92181$ is substituted.  We find
the critical temperature $\Tc$ is an increasing function of the binding 
energy $\eb$ near the unitarity limit.  The next-to-leading-order
correction is reasonably small compared to the leading term even at
$\eps=1$.  The naive extrapolation of the critical temperature to the
physical case of $d=3$ gives $\Tc/\eF\approx0.249$ in the unitarity
limit $\eb=0$. %0.248744 
This value is surprisingly close to results from two Monte Carlo
simulations, $\Tc/\eF=0.23(2)$~\cite{bulgac-Tc} and
$\Tc/\eF\approx0.25$~\cite{Akkineni-Tc}, while other two simulations
provide smaller values, $\Tc/\eF<0.14$~\cite{Lee-Schafer} and 
$\Tc/\eF=0.152(7)$~\cite{burovski-Tc}. 

It is also interesting to compare the critical temperature in the
unitarity limit with that in the BEC limit $T_\mathrm{BEC}$.  In the BEC
limit, all fermion pairs are confined into tightly bound molecules and
the system becomes a noninteracting Bose gas where the boson mass is
$2m$ and the boson density is $N/2$.  The critical temperature for the
Bose-Einstein condensation of such an ideal Bose gas at $d>2$ spatial
dimensions becomes
\begin{equation}
  \frac{T_\mathrm{BEC}}\eF
  =\frac12\left[\zeta\!\left(\frac d2\right)
  \Gamma\!\left(1+\frac d2\right)\right]^{-2/d}.
%  &=0.275664 - 0.0167191\,\eps + O(\eps^2). 
\end{equation}
To the leading and next-to-leading orders in $\eps=4-d$, the ratio of
the critical temperatures in the unitarity limit $\Tc$ and in the BEC
limit $T_\mathrm{BEC}$ at the same fermion density is given by
\begin{equation}
 \begin{split}
  \frac{\Tc}{T_\mathrm{BEC}}
  &=\sqrt{\frac89}\left[1+0.0177\,\eps+O(\eps^2)\right]\\  % 0.0177305
  &=0.943 + 0.0167\,\eps + O(\eps^2).
  \phantom{\frac{\int}{}}  % 0.942809 + 0.0167165
 \end{split}
\end{equation}
The ratio is slightly below unity, indicating the lower critical
temperature in the unitarity limit $\Tc<T_\mathrm{BEC}$.  The leading
order term of the above ratio, $\Tc/T_\mathrm{BEC}=\sqrt{8/9}$, has the
following clear physical interpretation: The critical temperature for
the Bose-Einstein condensation at $d=4$ is proportional to a square root
of the boson's density.  In the BEC limit, all fermion pairs form the
bound bosons, while only 8 of 9 fermion pairs form the bosons and 1 of 9
fermion pairs is dissociated in the unitarity limit [see the leading
order terms of fermion and boson densities in Eqs.~(\ref{eq:NF}) and
(\ref{eq:NB_regular})].  Thus their ratio in the critical temperature
should be $\Tc/T_\mathrm{BEC}=\sqrt{8/9}<1$ at $d=4$.

The more appropriate estimate of $\Tc$ at $d=3$ will be obtained by
matching the $\eps$ expansion around four spatial dimensions with the
exact result around $d=2$.  The critical temperature at unitarity in the
expansion over $\bar\eps=d-2$ is given by
$\Tc=\left(e^\gamma/\pi\right)\Delta$, where 
$\Delta/\eF=\left(2/e\right)e^{-1/\bar\eps}$ is the energy gap of the
fermion quasiparticle at zero temperature~\cite{Nishida-Son2}: 
\begin{equation}\label{eq:Tc-2d}
 \frac\Tc\eF = \frac{2e^{\gamma-1}}{\pi}
  \,e^{-1/\bar\eps}\left[1+O(\bar\eps)\right].
\end{equation}
We shall write the power series of $\bar\eps$ in the form of the Borel
transformation, 
\begin{equation}\label{eq:borel}
 \frac{\Tc(\bar\eps)}\eF = \frac{2e^{\gamma-1}}{\pi}\,e^{-1/\bar\eps} 
  \int_0^\infty\!dt\, e^{-t}B_\Tc(\bar\eps t).
\end{equation}
$B_\Tc(t)$ is the Borel transform of the power series in
$\Tc(\bar\eps)$, whose Taylor coefficients at origin is given by
$B_\Tc(t)=1+\cdots$.  In order to perform the integration over $t$ in
Eq.~(\ref{eq:borel}), the analytic continuation of the Borel transform 
$B_\Tc(t)$ to the real positive axis of $t$ is necessary.  Here we
employ the Pad\'e approximant, where $B_\Tc(t)$ is replaced by the
following rational functions 
\begin{equation}
 B_\Tc(t) = \frac{1+p_1 t+\cdots+p_M t^M}{1+q_1 t+\cdots+q_N t^N}\,.
\end{equation}
Then we incorporate the results around four spatial dimensions in
Eq.~(\ref{eq:Tc}) by imposing 
\begin{equation}\label{eq:boundary}
 \frac{\Tc(2-\eps)}\eF=0.260-0.0112\,\eps+\cdots
\end{equation}
on the Pad\'e approximants as a boundary condition. 
Since we have two known coefficients from the $\eps$ expansion, the
Pad\'e approximants $[M/N]$ satisfying $M+N=2$ are possible. 
Since we could not find a solution satisfying the boundary condition 
in Eq.~(\ref{eq:boundary}) for $[M/N]=[1/1]$, we adopt other two Pad\'e
approximants with $[M/N]=[2/0],\,[0/2]$, whose coefficients $p_m$ and
$q_n$ are determined uniquely by the above conditions.

\begin{figure}[tp]
 \includegraphics[width=0.45\textwidth,clip]{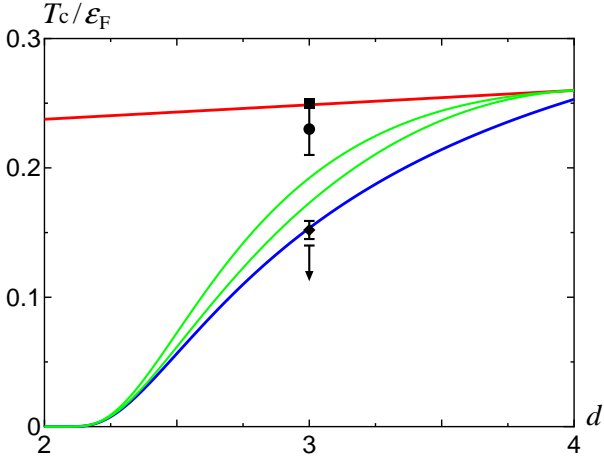}
 \caption{(Color online) The critical temperature $\Tc$ at unitarity as
 a function of the spatial dimension $d$.  The upper solid line is from
 the expansion around $d=4$ in Eq.~(\ref{eq:Tc}), while the lower solid
 curve is from the expansion around $d=2$ in Eq.~(\ref{eq:Tc-2d}).  The
 middle two curves show the different Borel-Pad\'e approximations
 connecting the two expansions.  The symbols at $d=3$ indicate the
 results from the Monte Carlo simulations;
 $\Tc/\eF=0.23(2)$~\cite{bulgac-Tc} (circle),
 $\Tc/\eF<0.14$~\cite{Lee-Schafer} (down arrow),
 $\Tc/\eF=0.152(7)$~\cite{burovski-Tc} (diamond), and
 $\Tc/\eF\approx0.25$~\cite{Akkineni-Tc} (square).  \label{fig:Tc}}
\end{figure}

Figure \ref{fig:Tc} shows the critical temperature $\Tc$ in units of the
Fermi energy $\eF$ as a function of the spatial dimension $d$.  The
middle two curves show $\Tc/\eF$ in the different Pad\'e approximants
connecting the two expansions around $d=4$ and $d=2$.  These
Borel-Pad\'e approximations give $\Tc/\eF=0.173$ and $0.192$ at $d=3$,
%0.172912, 0.192202
which are located between the naive extrapolations to $d=3$ from the 
$\eps=4-d$ expansion $(\Tc/\eF\to0.249)$ and the $\bar\eps=d-2$
expansion $(\Tc/\eF\to0.153)$. %0.153452
It is also interesting to compare our results with those from the recent
Monte Carlo simulations, where $\Tc/\eF=0.23(2)$~\cite{bulgac-Tc},
$\Tc/\eF<0.14$~\cite{Lee-Schafer}, $\Tc/\eF=0.152(7)$~\cite{burovski-Tc},
and $\Tc/\eF\approx0.25$~\cite{Akkineni-Tc}.  Although these results
from the Monte Carlo simulations seem not to be settled, the
interpolation of the two expansions provides the moderate value 
$\Tc/\eF=0.183\pm0.014$ not too far from the Monte Carlo simulations.
%0.182557 \pm0.0136398 

\subsection{Thermodynamic functions at $\Tc$}
Finally we show the thermodynamic functions at $\Tc$ in the unitarity
limit $\eb=0$ to the leading and next-to-leading orders in $\eps$.  The
pressure $P$ normalized by the fermion density $\eF N$ follows from
Eqs.~(\ref{eq:P-Tc}), (\ref{eq:N-Tc}), and (\ref{eq:eF-Tc}). Introducing
the numerical values $C_P\approx8.4144$ and $C_N\approx1.92181$, we
obtain the pressure up to the next-to-leading order in $\eps$ as
\begin{equation}
 \left.\frac P{\eF N}\right|_{\Tc} = 0.116+0.0188\,\eps. 
  %0.116065 +0.0187835
\end{equation}
From the universal relationship in the unitarity limit $E=(d/2)P$, the 
energy density is given by
\begin{equation}
 \left.\frac E{\eF N}\right|_{\Tc} = 0.232-0.0205\,\eps.
  %0.23213 -0.0204654
\end{equation}
The chemical potential at the critical temperature $\mu=\eps\,\Tc\ln2$
is $O(\eps)$.  Normalizing $\mu$ by the Fermi energy in
Eq.~(\ref{eq:Tc}), we have
\begin{equation}
 \left.\frac\mu\eF\right|_{\Tc} = \epsilon\ln2\sqrt{\frac2{3\pi^2}} 
  = 0.180\,\eps. %0.180148
\end{equation}
Then the entropy density $\Tc S=(d/2+1)P-\mu N$ is given by
\begin{equation}
 \left.\frac S{N}\right|_{\Tc} = 1.340 - 0.642\,\eps.
  %1.33973 -0.642118
\end{equation}
The next-to-leading-order corrections in the pressure and energy density
are reasonably small compared to the leading order terms, while that is
large for the entropy density.

We match the thermodynamic functions at $\Tc$ in the expansions over
$\eps=4-d$ with those around $d=2$ as we demonstrated for $\Tc/\eF$. 
The critical temperature around $d=2$ is $\Tc/\eF\sim e^{-1/\bar\eps}$, 
which is exponentially small and negligible compared to any power series
of $\bar\eps$.  Therefore, the pressure, the energy density, and the
chemical potential at $\Tc$ in the expansions over $\bar\eps=d-2$ are
simply given by those at zero temperature~\cite{Nishida-Son2}:
\begin{align}
 \left.\frac{P}{\eF N}\right|_{\Tc} 
 &= \frac{2}{d+2}\,\xi = \frac12-\frac58\bar\eps,\\
 \left.\frac{E}{\eF N}\right|_{\Tc}
 &= \frac{d}{d+2}\,\xi = \frac12-\frac38\bar\eps,
 \phantom{\frac{\frac\int\int}{\frac\int\int}}\hspace{-4.5mm} \\ 
 \left.\frac{\mu}{\eF}\right|_{\Tc} &= \xi = 1-\bar\eps.
\end{align}
A straightforward calculation shows that the entropy per particle at
$\Tc$ to the leading order in $\bar\eps$ is given by 
\begin{equation}
 \left.\frac S{N}\right|_{\Tc} = \frac{\pi^2}3\frac\Tc\eF
  = \frac{2\pi e^{\gamma-1}}{3}\,e^{-1/\bar\eps}.
\end{equation}

Using the Borel-Pad\'e approximations in order to connect the two
expansions above, the thermodynamic functions at $d=3$ are found to be
$\left.P/(\eF N)\right|_\Tc=0.172\pm0.022$, %0.172069 \pm0.0220483
$\left.E/(\eF N)\right|_\Tc=0.270\pm0.004$, %0.270048 \pm0.00402932
$\left.\mu/\eF\right|_\Tc=0.294\pm0.013$, %0.294172 \pm0.0131934
and $\left.S/N\right|_\Tc=0.642$. %0.642492
The errors here indicates only the uncertainty due to the choice of
different Pad\'e approximants. 
In the recent Monte Carlo simulation, the thermodynamic functions at 
the critical temperature is $\left.P/(\eF N)\right|_\Tc=0.207(7)$,
$\left.E/(\eF N)\right|_\Tc=0.31(1)$,
$\left.\mu/\eF\right|_\Tc=0.493(14)$, and
$\left.S/N\right|_\Tc=0.16(2)$~\cite{burovski-Tc}. 
We see that the interpolations of the two expansions indeed improve the 
series summations compared to the naive extrapolations from $d=4$ or 
$d=2$, while there still exist deviations between our results and the
Monte Carlo simulation.  We can understand these deviations partially 
due to the difference in the determined critical temperatures.  The
large deviations existing in $\mu/\eF$ and $S/N$ may be because we know
only the leading term for $\mu$ and the next-to-leading-order correction
to $S$ is sizable.

\section{Summary and concluding remarks \label{sec:summary}}
In this paper, the thermodynamics of the Fermi gas near the unitarity
limit at finite temperature has been investigated using the systematic 
expansion over $\eps=4-d$.  We discussed that the thermodynamics in the
low temperature region $T\ll\Tc$ is dominated by the bosonic phonon
excitations.  The analytic formulas for the thermodynamic functions at
the fixed fermion density are derived in the two limiting cases;
$T\ll\eps\Tc$ in Eqs.~(\ref{eq:mu-N1})--(\ref{eq:S-N1}) and 
$\eps\Tc\ll T\ll\Tc$ in Eqs.~(\ref{eq:mu-N2})--(\ref{eq:S-N2}). 

In the high temperature region $T\sim\Tc$, the fermionic quasiparticles
are excited as well as the bosonic quasiparticles.  We showed that the
similar power counting rule of $\eps$ to that developed at zero
temperature works even above $\Tc$.  The superfluid phase transition at
$T=\Tc$ is of the second order.  The critical temperature $\Tc$ and the
thermodynamic functions around $\Tc$ were calculated to the leading and
next-to-leading orders in $\eps$.  We found the critical temperature is
an increasing function of the binding energy $\eb$ near the unitarity
limit:
\begin{equation}
 \frac\Tc\eF = 0.260-0.0112\,\eps+0.0234\,\frac\eb\eF.
  %0.259899 -0.0111548 +0.0234102
\end{equation}
The next-to-leading-order correction is reasonably small compared to the
leading term even at $\eps=1$.  In the unitarity limit $\eb=0$, the
naive extrapolation of the critical temperature to the physical case of
$d=3$ gives $\Tc/\eF\approx0.249$. %0.248744 

We also discussed the matching of the $\eps$ expansion around $d=4$ with
the expansion around $d=2$.  The critical temperature at unitarity in the
expansion over $\bar\eps=d-2$ is given by
$\Tc/\eF=\left(2e^{\gamma-1}/\pi\right)e^{-1/\bar\eps}$, where its naive 
extrapolation to $\bar\eps=1$ gives $\Tc/\eF\approx0.153$. 
The Borel-Pad\'e approximations connecting the two expansions yielded
$\Tc/\eF=0.183\pm0.014$ at $d=3$, which is a moderate value located
between the two naive extrapolations (Fig.~\ref{fig:Tc}).  These values
are not too far from the results obtained by the recent Monte Carlo
simulations where
$\Tc/\eF=0.15$--$0.25$~\cite{bulgac-Tc,burovski-Tc,Akkineni-Tc}. 
We also applied the Borel-Pad\'e approximations to the thermodynamic
functions at $\Tc$, which yielded 
$\left.P/(\eF N)\right|_\Tc\approx0.172$, 
$\left.E/(\eF N)\right|_\Tc\approx0.270$,
$\left.\mu/\eF\right|_\Tc\approx0.294$,  
and $\left.S/N\right|_\Tc\approx0.642$ at $d=3$. 

The Borel-Pad\'e approximations employed here to match the two
expansions around four and two spatial dimensions do not correctly 
reflect large-order behaviors of the two expansions, i.e., the
expansions over $\eps=4-d$ and $\bar\eps=d-2$ are most probably not
convergent and hence the Borel transform $B(t)$ of such series
expansions will have singularities somewhere in the complex
$t$-plane~\cite{Nishida-Son2}. 
In order for the accurate determination of the critical temperature and
the thermodynamic functions at $d=3$, it will be important to
appropriately take into account the knowledge on the large-order
behaviors of the expansions both around four and two spatial
dimensions.  The calculation of higher-order corrections to our results
is also interesting for this purpose.  These problems should be studied
in future works.

\begin{acknowledgments}
 The author would like to thank D.~T.~Son for useful discussions.  
 This research was supported by the Japan Society for the Promotion of 
 Science for Young Scientists. 
\end{acknowledgments}

\end{document}